\documentclass[letterpaper, 10 pt, conference]{ieeeconf}  % Comment this line out if you need a4paper

\IEEEoverridecommandlockouts                              % This command is only needed if 
\usepackage{graphics} % for pdf, bitmapped graphics files
\usepackage{epsfig} % for postscript graphics files
\usepackage{mathptmx} % assumes new font selection scheme installed
\usepackage{times} % assumes new font selection scheme installed
\usepackage{amsmath} % assumes amsmath package installed
\usepackage{amssymb}  % assumes amsmath package installed

\usepackage{prettyref}  % assumes amsmath package installed
\usepackage{graphicx}
\usepackage{cite}
\usepackage{caption}
\usepackage{subcaption}
\usepackage{prettyref}
\newrefformat{fig}{Fig.~\ref{#1}}
\newrefformat{table}{Table~\ref{#1}}
\newrefformat{eq}{Eq.~(\ref{#1})}

\usepackage[english]{babel}
\usepackage{hyperref}

\title{\LARGE \bf
Turbulence-based load alleviation control for wind turbine in extreme turbulence situation}

\author{Liang Dong$^{1}$ and Wai Hou Lio$^{1}$% <-this % stops a space
\thanks{*This research was supported by the ForskEL Programme under project PowerKey - Enhanced wind turbine control for optimised wind power plant operation (Grant No. 12558) and the EUDP LICOREIM - LIdar-assited COntrol for REliability IMprovement (Grant No. 64019-0580).}% <-this % stops a space
\thanks{$^{1}$Department of Wind Energy, Technical University of Denmark (DTU), Frederiksborgvej 399, 4000 Roskilde, Denmark.
{\tt\small ldong@dtu.dk};
{\tt\small wali@dtu.dk}}%
}

\begin{document}

\maketitle
\thispagestyle{empty}
\pagestyle{empty}

%%%%%%%%%%%%%%%%%%%%%%%%%%%%%%%%%%%%%%%%%%%%%%%%%%%%%%%%%%%%%%%%%%%%%%%%%%%%%%%%
\begin{abstract}
The extreme loads experienced by the wind turbine in the extreme wind events are critical for the evaluation of structural reliability. Hence, the load alleviation control methods need to be designed and deployed to reduce the adverse effects of extreme wind events. This work demonstrates that the extreme loads are highly correlated to wind conditions such as turbulence-induced wind shears. Based on this insight, this work proposes a turbulence-based load alleviation control strategy for adapting the controller to changes in wind condition. The estimation of the rotor averaged wind shear based on the rotor loads is illustrated, and is herein used to statistically characterize the extreme wind events for control purpose. To demonstrates the benefits, simulations are carried out using high-fidelity aero-elastic tool and the DTU 10 MW reference turbine in normal and extreme turbulence wind conditions. The results indicate that the proposed method can effectively decrease the exceedance probability of the extreme loads. Meanwhile, the method can minimize the loss of annual energy production in normal operating condition.

\end{abstract}

\section{INTRODUCTION}

With the advent of larger rotor size and more flexible wind turbine, limiting the loads experienced in extreme turbulence condition is becoming increasingly important. The wind energy industry is continuously researching better control methods to achieve a reasonable trade-off between energy production and component loading \cite{Loew2020,Odgaard2016}. The probability of extreme blade and tower cyclic stresses are highly correlated with the wind condition. The power production cases in extreme turbulence are among the top design driving load cases. Environmental wind conditions often change with time, for example, wake in a wind farm, extreme shear in complex terrain, high turbulence weather fronts, etc. Therefore, an optimal trade-off can be achieved in principle if the controller can adapt its behaviour to various wind conditions. 

Turbulence Intensity (TI) is commonly used as an indicator of spatial-temporal variation in wind inflow. To achieve the real-time knowledge of the wind, the concept of wind speed estimation based on the wind turbine operating condition has been proposed in a considerable amount of literature \cite{Gocmen2016,Lio2021,Jena2015,Kanev2017,DIckler2019RequirementsControl}. In addition, another growing body of literature \cite{Schlipf2020,Pena2017} investigated on estimating turbulence intensity using Light Detection And Ranging (LiDAR) systems. The work \cite{Schlipf2020} proposed a solution of scheduling a feedback controller based on the LiDAR data to reduce the structural loads. The work \cite{Abdallah2016InfluenceReliability} discussed the improving structural reliability based on advanced control algorithm in extreme wind situation, which yield a reduction in the extreme load distribution.

Most of the existing studies have focused on effective wind speed based control, but the spatial imbalance in wind inflow is not included explicitly, which has a significant impact on generating extreme load for the large sized wind turbines. The turbulent kinetic energy within the rotor disk will produce the imbalanced local wind, then it can be transported into the blade structure, finally the load will be propagated into the hub and other components. Several attempts have been made to estimate the rotor averaged shear information \cite{Simley2016,Bertele2017WindHarmonics}, which can be integrated into controller as an indicator of the extreme  wind events.

The aim of this work is to explore the relationship between the loads and statistical quantities of rotor averaged wind condition. Then, an algorithm for modifying the control parameters in response to the indicators of turbulence-induced shear is proposed. The method is referred to as Turbulence-based Load Alleviation Control (TLAC).

The remainder of this paper is organized as follows. \prettyref{sec:SimuSetup} briefly describes the simulation setup used in this work. \prettyref{sec:RAWCEstimation} deals with the estimation of the rotor averaged wind condition. The correlation between loads and wind condition is analyzed in \prettyref{sec:RelationshipLoadTurbulence}. In \prettyref{sec:TBC}, the TLAC framework is developed and the simulation results are presented, comparing the TLAC with the baseline control. The conclusion is summarized in \prettyref{sec:conclusion}.

\section{Simulation Setup}\label{sec:SimuSetup}
To evaluate the performance of TLAC, the simulations are performed using the aero-elastic tool HAWC2 and the DTU 10 MW reference wind turbine~\cite{Bak2013} with a diameter of 178.3 m and a hub height of 119 m. The rated wind speed is 11.4 m/s. The DTU Wind Energy Controller (DTUWEC)~\cite{Meng2020DTUWEC:Features} is referred to as baseline controller in this work, which is an open-source and conventional variable speed controller. The DTUWEC is able to couple with aero-elastic simulation code to investigate the performance of various control strategy.

\subsection{Coordinate System}
\begin{figure}
	\centering
	\includegraphics[width=8cm,trim={0cm 0.5cm 0 0cm},clip]{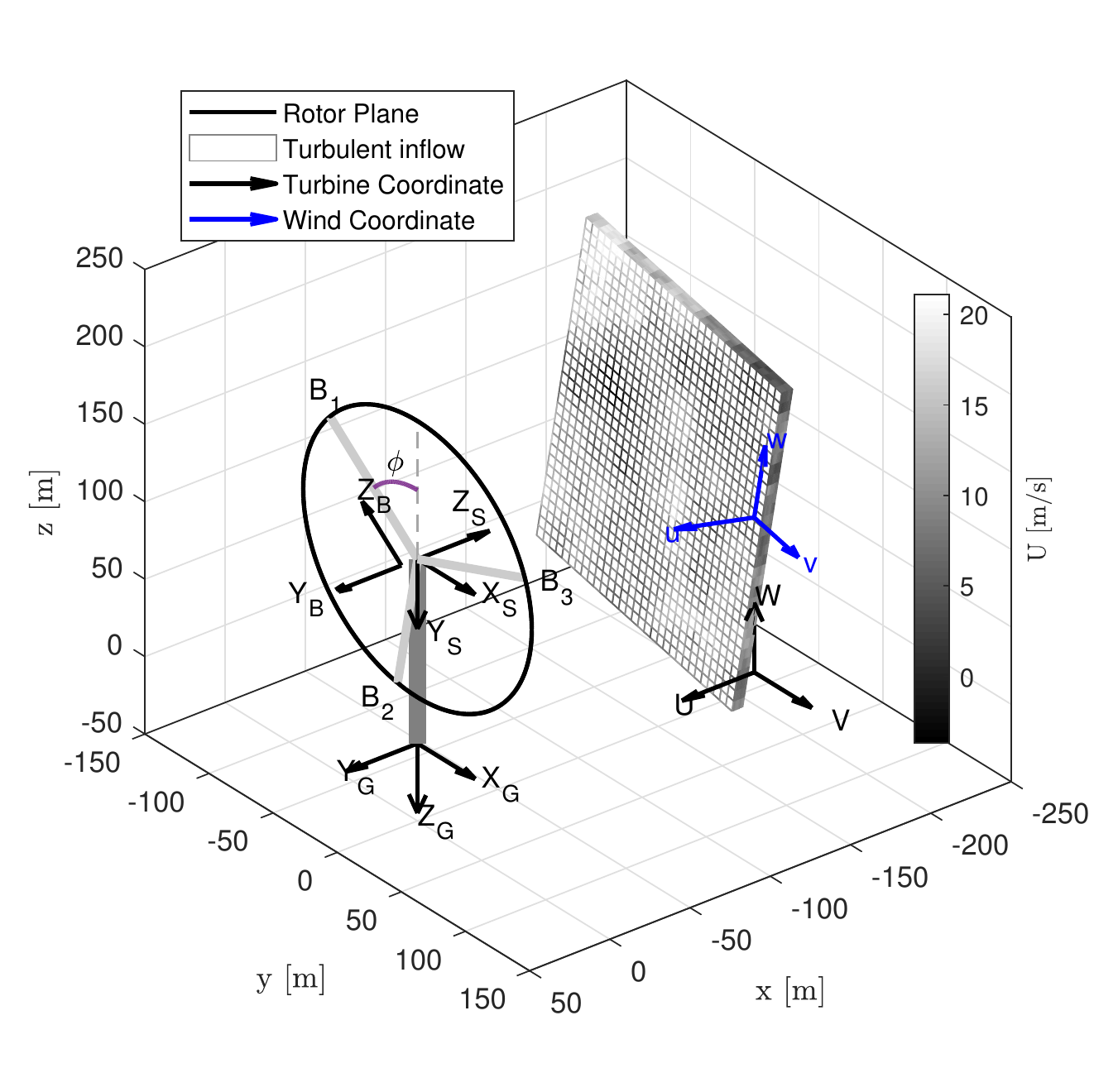}
	\caption{Illustration of right-handed coordinates. X-Y-Z represents coordinate for each turbine component, the subscripts 'G', 'B', 'S' stand for tower base (ground), blade and stationary shaft, respectively. U-V-W represents the original wind coordinate without rotation.}
	\label{fig:CoordinateSystem}
\end{figure}

The coordinate systems for each major component are illustrated in \prettyref{fig:CoordinateSystem}, including the tower, blade and shaft coordinates. When viewing the turbine from upstream, clockwise azimuth angle $\varphi$ = 0 radian represents blade 1 pointing upward, the azimuth angle of blade 2 and 3 are $\varphi+2\pi/3$ and $\varphi+4\pi/3$ respectively.

The wind field described in wind coordinate system $(u,v,w)$ is rotated horizontally and vertically with respect to the coordinate system $(U,V,W)$. The positive horizontal and vertical direction are pointing to left and upwards when viewing the turbine from the upwind. 
%The wind speed in the wind coordinate system is transformed into rotor coordinate system as follows:
%\begin{eqnarray}
%\left[\begin{matrix}U\\V\\W\\\end{matrix}\right] = &&
%\left[\begin{matrix}\cos\left(\alpha_h\right)&-\sin\left(\alpha_h\right)&0\\\sin\left(\alpha_h\right)&\cos\left(\alpha_h\right)&0\\0&0&1\\\end{matrix}\right]\nonumber\\&&\times
%\left[\begin{matrix}\cos\left(\alpha_v\right)&0&-\sin\left(\alpha_v\right)\\0&1&0\\\sin\left(\alpha_v\right)&0&\cos\left(\alpha_v\right)\\\end{matrix}\right]
%\left[\begin{matrix}u\\v\\w\\\end{matrix}\right]
%\end{eqnarray}
%where $\alpha_v$ is vertical inflow angle, $\alpha_h$ is horizontal inflow angle.

\subsection{Load Case Definition}\label{sec:DLC}
According to the IEC standard \cite{InternationalElectrotechnicalCommissionandothers2019WindRequirements}, two subsets of the normal power producing Design Load Cases (DLC) have been evaluated, including Normal Turbulence Model (NTM) DLC 1.2 and Extreme Turbulence Model (ETM) DLC 1.3. Since the peak loads often occur around rated wind speed, 6 random seeds at each mean wind speed from 10 to 14~m/s spaced at 2~m/s apart are simulated. The turbulence category is A. For each mean wind speed, three different wind directions 8, 0 and -8 deg are included. The simulation time is 700 s and the first 100 s is discarded. Finally, in total 36 cases are obtained, including 18 ETM cases and 18 NTM cases.

\section{Rotor Averaged Wind Condition Estimation}\label{sec:RAWCEstimation}
This section describes the method to estimate the rotor averaged wind characteristics based on the rotor loads.

\subsection{Actual Rotor Averaged Wind Characteristic}
The rotor averaged wind speed and shear are modelled as the least-squares fit of the $u$ component wind speeds across the rotor disk area:
\begin{equation}
\left[\begin{matrix}u_1\\u_2\\\begin{matrix}\vdots\\u_N\\\end{matrix}\\\end{matrix}\right]=
\left[\begin{matrix}1&\Delta{y}_1&\Delta{z}_1\\1&\Delta{y}_2&\Delta{z}_2\\\vdots&\vdots&\vdots\\1&\Delta{y}_N&\Delta{z}_N\\\end{matrix}\right]
\left[\begin{matrix}U_\mathrm{eff}\\\delta_h\\\delta_v\\\end{matrix}\right].
\end{equation}
where for all $N$ points on the rotor disk, $U_\mathrm{eff}$ is the rotor averaged wind speed. The linear variation of wind speed across the rotor disc is represented by rotor averaged vertical shear $\delta_v$ and horizontal shear $\delta_h$, $\Delta{y}$ and $\Delta{z}$ are the horizontal and vertical distance between grid points and rotor centre.

The rotor averaged turbulence intensity $I_\mathrm{eff}$ and resultant shear magnitude $\delta$ are defined as
\begin{equation}
%\begin{aligned}
    I_\mathrm{eff} = \frac{{\sigma_U}_\mathrm{eff}}{\bar{U}_\mathrm{eff}},\qquad 
    \delta = \sqrt{\delta_h^2 + \delta_v^2} .\\
%\end{aligned}
\end{equation}
where $\bar{U}_\mathrm{eff}$ denotes the mean value and ${\sigma_U}_\mathrm{eff}$ is the standard deviation of the rotor averaged wind speed.

\subsection{Wind Condition Estimation}\label{sec:WSE}
Torque balance method is commonly utilised to estimate the rotor averaged wind speed since it does not require additional sensors. The basic measurements of the generator reaction torque $Q_g$, the rotor speed $\Omega$ and the pitch angle $\beta$ are sufficient~\cite{HouLio2020EffectiveDown-regulation}. 
In addition, the rotor averaged wind speed $U_\mathrm{eff}$ can also be estimated according to the thrust balance equation, in which additional blade root load sensors are required:
\begin{equation}
F_T=\frac{1}{2}C_t\left(\beta,\lambda\right)\rho{U^2_\mathrm{eff}}A_d.
\end{equation}
where $C_t\left(\beta,\lambda\right)$ denotes the thrust coefficient that is a function of the pitch angle $\beta$ and the tip-speed ratio $\lambda$, $\rho$ and $A_d$ represent the air density and rotor disk area, respectively.

The thrust is the dominant source of each blade root out-of-plane (oop) bending moment $M_{\mathrm{oop},i}$:
\begin{equation}
M_{\mathrm{oop},i}=\frac{1}{2N_b}C_t\left(\beta,\lambda\right)\rho{U^2_{\mathrm{b},i}}A_dR_\mathrm{eq},
\label{eq:ThrustBalance}
\end{equation}
where $N_b$ denotes the number of blades, $U_{\mathrm{b},i}$ is blade-equivalent wind speed and $R_\mathrm{eq}$ is the equivalent radius. The contribution of the gravity, inertia loading are neglected in \prettyref{eq:ThrustBalance} due to the aerodynamic loading is more significant. Meanwhile, the impact of yaw misalignment and inflow angle are not included. The moment $M_{\mathrm{oop},i}$ is considered as an integral result of the thrust force along the blade span-wise direction. Hence, the equivalent radius $R_\mathrm{eq}$ is determined by assuming the single equivalent concentrated loads applied to that position, where it produces approximately the equivalent aerodynamic loads as the actual span-wise distribution of thrust. The equivalent radius $R_\mathrm{eq}$ is defined based on the steady-state response:
\begin{equation}
R_\mathrm{eq}=\frac{M_\mathrm{oop}(\bar{U})}{\int_{0}^{R_\mathrm{b}}F_\mathrm{oop}\left(r,\bar{U}\right)\mathrm{d}r}.
\end{equation}
where $\bar{U}$ denotes mean wind speed, $R_\mathrm{b}$ is rotor radius, $M_\mathrm{oop}\left(\bar{U}\right)$ is the blade root out-of-plane  bending moment, $F_\mathrm{oop}\left(r,\bar{U}\right)$ is the the distributed out-of-plane force along the blade span-wise direction $r$, which can be calculated by the steady-state response of the turbine using aero-elastic analysis code HAWCStab2 \cite{Hansen2018}.

From the blade load measurement, the blade-equivalent wind speed $U_{\mathrm{b},i}$ can be estimated according to \prettyref{eq:ThrustBalance}, which can also be expressed as the combination of the wind speed and shear:
\begin{equation}
\begin{aligned}
U_{\mathrm{b},i} &=U_\mathrm{eff} - \delta_hR_\mathrm{eq}\sin(\varphi+\frac{2\pi}{3}(i-1)) \\
&\quad +  \delta_vR_\mathrm{eq}\cos(\varphi+\frac{4\pi}{3}(i-1)).
\end{aligned}
\end{equation}

Subsequently, the rotor averaged wind characteristics are derived by transforming $U_{\mathrm{b},i}$ into non-rotating coordinate:
\begin{eqnarray}
\left[\begin{matrix}
\hat{U}_\mathrm{eff}, \hat{\delta}_v, \hat{\delta}_h\end{matrix}\right]^\mathrm{T} = 
\mathbf{Q}\left[\begin{matrix}U_{\mathrm{b},1}, U_{\mathrm{b},2}, U_{\mathrm{b},3}\end{matrix}\right]^\mathrm{T},
\end{eqnarray}
where the symbol $\hat{(.)}$ denotes the estimation results, the superscript $(.)^\mathrm{T}$ denotes the transpose of a matrix, the matrix $\mathbf{Q}$ is expressed as:
\begin{eqnarray}
\mathbf{Q} = \frac{2}{3R_\mathrm{eq}}\left[\begin{matrix}\frac{R_\mathrm{eq}}{2}&\frac{R_\mathrm{eq}}{2}&\frac{R_\mathrm{eq}}{2}\\\cos\left(\varphi\right)&\cos\left(\varphi+\frac{2\pi}{3}\right)&\cos\left(\varphi+\frac{4\pi}{3}\right)\\-\sin\left(\varphi\right)&-\sin\left(\varphi+\frac{2\pi}{3}\right)&-\sin\left(\varphi+\frac{4\pi}{3}\right)\\
\end{matrix}\right].
\end{eqnarray}

%\begin{figure}
%\centering
%	\newline
%    \begin{subfigure}{0.49\textwidth}
%	\centering
%	\includegraphics[width=7.5cm]{BladeForceDistribution2.eps}
%	\caption{Normalized blade root out-of-plane force distribution.}
%	\label{fig:BladeForceDistribution}
%    \end{subfigure}
%	\newline
%	 \begin{subfigure}{0.49\textwidth}
%	\centering
%	\includegraphics[width=8cm]{BladeEqLength.pdf}
%	\caption{The ratio between blade equivalent length and rotor radius in various mean wind speed.}
%	\label{fig:BladeEqLength}
%    \end{subfigure}
%	\newline
%	\caption{Load distribution and blade equivalent length of DTU 10MW reference turbine.}
%	\label{fig:DTU10MWDistribution}
%\end{figure}

\subsection{Validation of Estimation Method}
\begin{figure}
\centering
%	\centering
	\includegraphics[width=8.6cm,trim={0cm 0cm 0 0cm},clip]{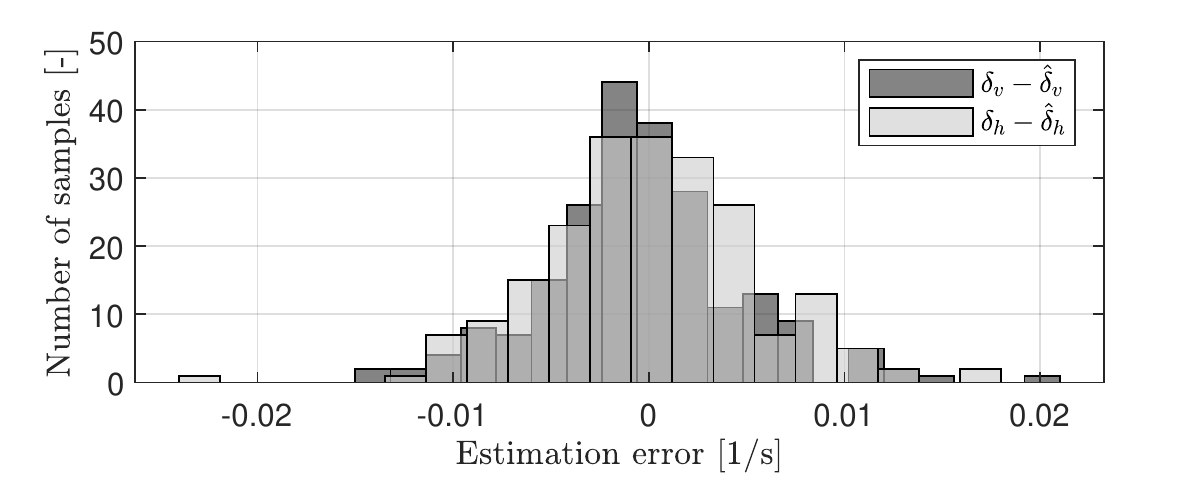}
	\caption{Histogram of the shear estimation errors.}
	\label{fig:EstimationError}
\end{figure}

To assess the accuracy of the proposed estimation method, the estimation procedure is carried out for all the DLCs described in \prettyref{sec:DLC}. The calibration is performed to remove the constant errors from the unmodeled dynamics as mentioned in \prettyref{sec:WSE}. All time series are divided into 6 segments, in total of 216 segments. The histogram of estimation errors for $\delta_v$ and $\delta_h$ is shown in \prettyref{fig:EstimationError}. Although there are some small deviations, the estimation accuracy is acceptable in general.
%\begin{figure}
%\centering
%	\centering
%	\includegraphics[width=7cm,trim={0cm 0cm 0cm 0cm},clip]{Timeseries_14s03_Wind.eps}
%	\caption{Comparison of estimation and actual wind with the ETM at a mean wind speed of 14m/s.}
%	\label{fig:Timeseries_14s03_Wind}
%\end{figure}

The magnitude-squared coherence is used to indicate how well estimation corresponds to actual wind condition at each frequency. The coherence $C_{xy}(f)$ is a function of the power spectral densities, $P_{xx}(f)$ and $P_{yy}(f)$, and the cross power spectral density, $P_{xy}(f)$, of the signals $x$ and $y$:
\begin{equation}
    C_{xy}(f) = \frac{\lvert{P_{xy}(f)\lvert^2}}{P_{xx}(f)P_{yy}(f)} .
\end{equation}

\begin{figure}
\centering
%	\centering
	\includegraphics[width=9cm,trim={0.5cm 0cm 0 0cm},clip]{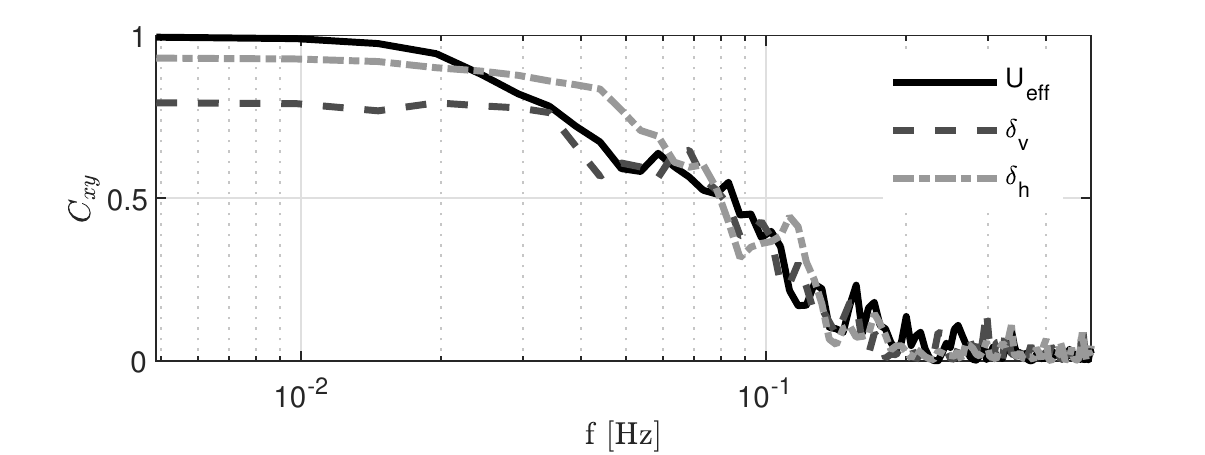}
	\caption{Magnitude-squared coherence between actual and estimated wind characteristics.}
	\label{fig:Coherence_WSE}
\end{figure}

The coherence results, including the $U_\mathrm{eff}$, $\delta_v$ and $\delta_h$, are shown in \prettyref{fig:Coherence_WSE}. The frequency at which the coherence equals 0.5 is around 0.07-0.09 Hz, indicating that the estimation can capture the general trend of actual value. The good correlation between estimation and actual wind provides the foundation for the turbulence-based control concept. The estimation results will be further discussed in \prettyref{sec:Simulation}.

\section{Relationship Between Load and Turbulence}\label{sec:RelationshipLoadTurbulence}

Three major components of wind turbine are considered in this work: tower bottom fore-aft bending moment $M_{x,\mathrm{t}}$, blade root out-of-plane bending moment $M_\mathrm{oop}$, hub tilt bending moment $M_{x,\mathrm{h}}$.
%\begin{enumerate}
%    \item Tower bottom fore-aft bending moment $M_{x,\mathrm{t}}$;
%    \item Blade root out-of-plane bending moment $M_\mathrm{oop}$;
%    \item Hub tilt bending moment $M_{x,\mathrm{h}}$.
%\end{enumerate}

\begin{figure}
\centering
%	\centering
	\includegraphics[width=8.5cm,trim={0.5cm 0cm 0 0cm},clip]{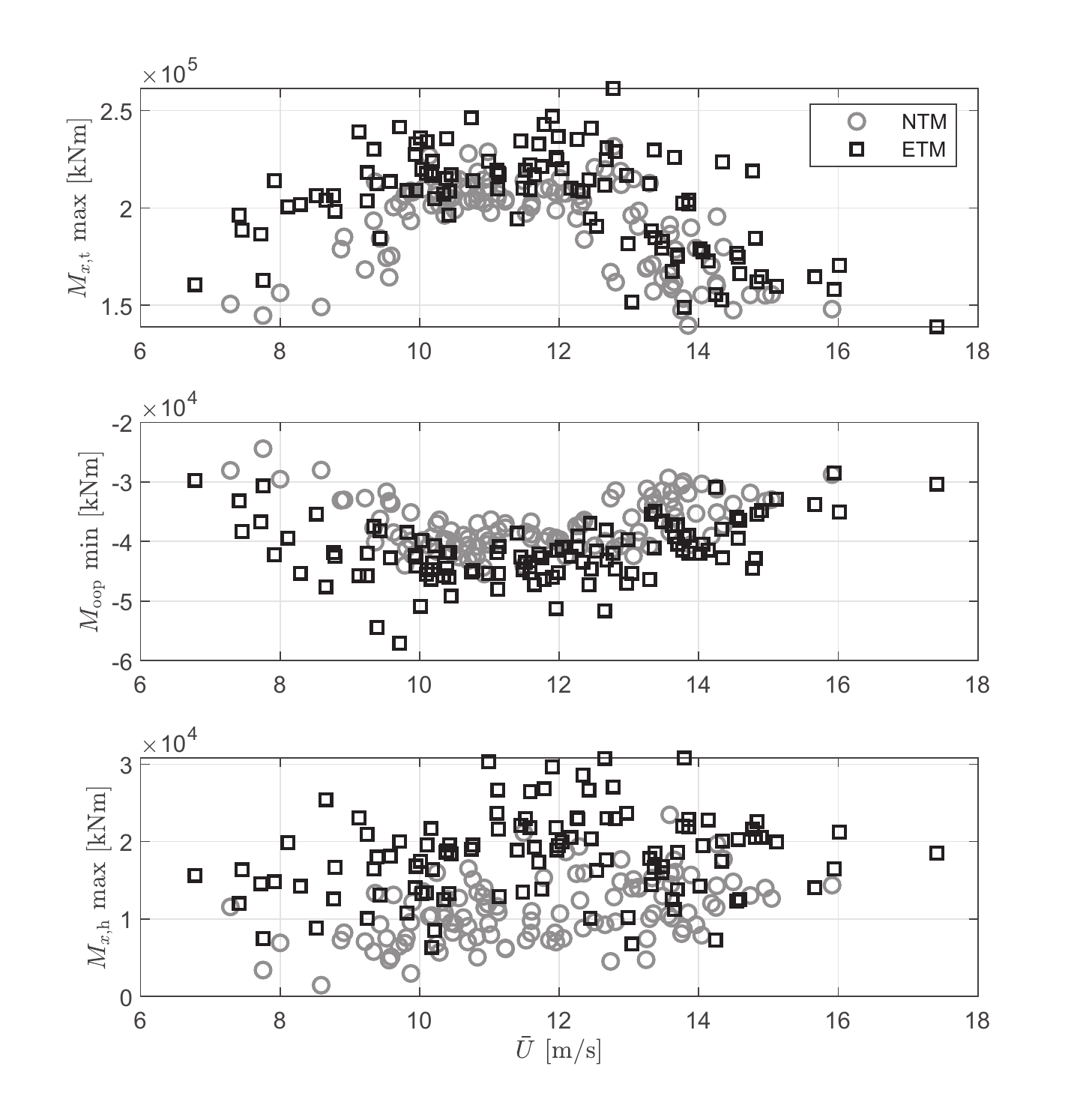}
	\caption{Relationship between extreme load and mean wind speed. From top to bottom: tower, blade,  hub. Notice that the sign of the blade root $M_\mathrm{oop}$ is negative.}
	\label{fig:ExLoad_WindSpd}
\end{figure}

To identify the relationship between loads and wind condition, the extreme load and wind condition are extracted from the previously divided 216 segments. The load distribution in \prettyref{fig:ExLoad_WindSpd} illustrates a clear trend in NTM and ETM loads. The ETM cases always produce larger extreme loads around rated wind speed. The extreme loads exist around the rated wind speed 11.4~m/s, hence in this work the wind speed range is focused on $U\in[U_\mathrm{LB}, U_\mathrm{UB}], U_\mathrm{LB} = 8 \mathrm{~m/s}, U_\mathrm{UB} = 16 \mathrm{~m/s}$
%\begin{equation}
%U\in[U_\mathrm{LB}, U_\mathrm{UB}], \qquad %U_\mathrm{LB} = 8 \mathrm{m/s}, \quad U_\mathrm{UB} %= 16 \mathrm{m/s}
%\end{equation}
where the subscripts '$\mathrm{LB}$' and '$\mathrm{UB}$' represent the lower and upper bound. 

\begin{figure}
\centering
%	\centering
	\includegraphics[width=8.5cm,trim={0.5cm 0cm 0 0cm},clip]{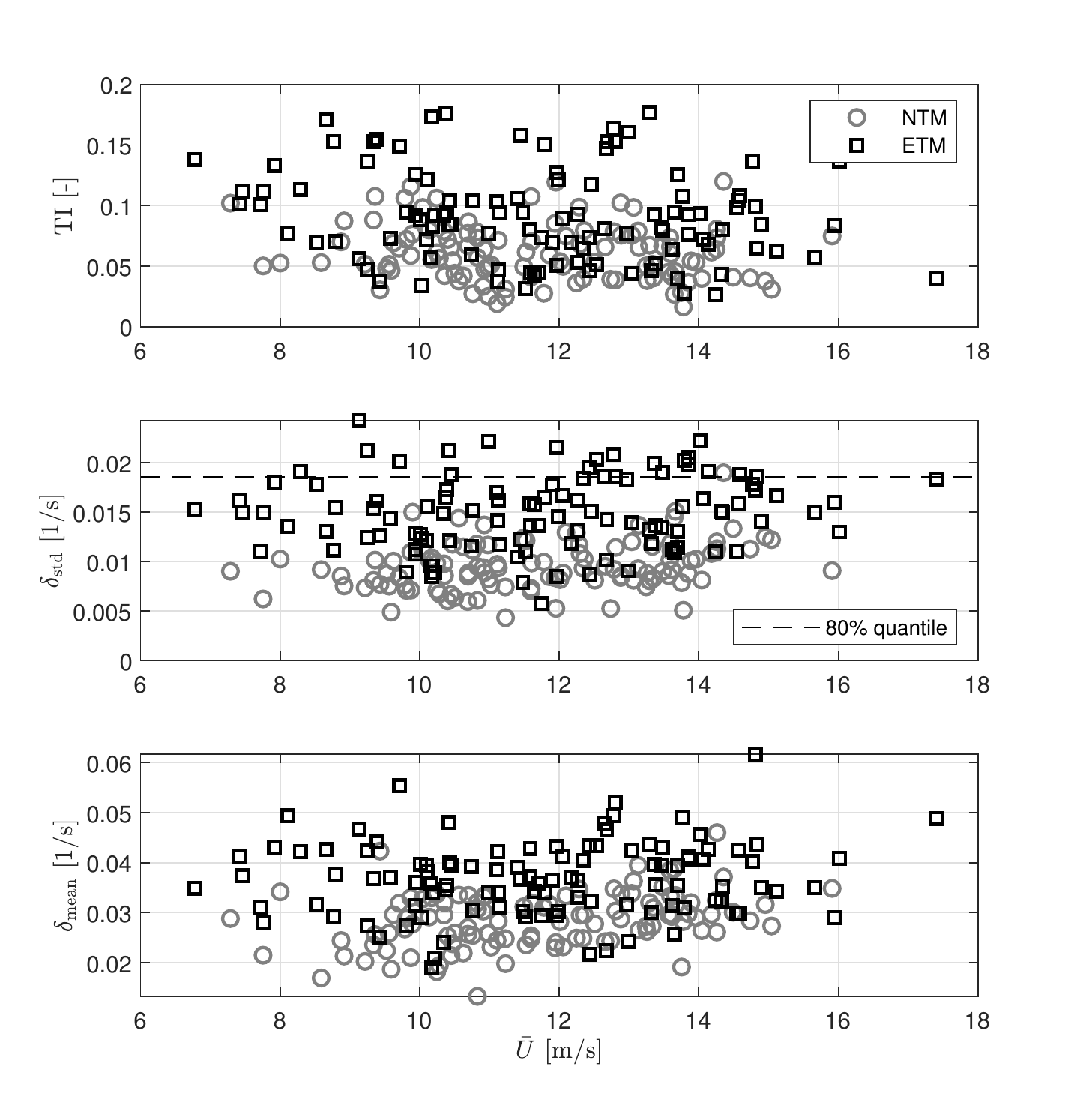}
	\caption{Statistical quantities in ETM and NTM cases. From top to bottom: rotor averaged TI, standard deviation of resultant shear, averaged value of resultant shear.}
	\label{fig:Relationship_ETMNTM}
\end{figure}

For assessing whether the estimated wind characteristics can be used to detect the extreme wind condition, \prettyref{fig:Relationship_ETMNTM} presents the statistical quantities including the rotor averaged turbulence intensity $I_\mathrm{eff}$, mean and standard deviation of the resultant shear $\delta_\mathrm{mean}$, $\delta_\mathrm{std}$. It is generally observed that the differences between ETM and NTM for the quantities $\delta_\mathrm{mean}$ and $\delta_\mathrm{std}$ are more obvious than that for  $I_\mathrm{eff}$, especially around the rated wind speed. It is somewhat surprising that $I_\mathrm{eff}$ in ETM is close to that in NTM, which reveals that it is not sufficient to detect extreme wind condition solely from $I_\mathrm{eff}$. The shear information would apparently be better indicators. The 80\% quantile is shown as the dashed line. ETM and NTM sets are to some extend separable by a threshold, which can be used for control purpose in \prettyref{sec:TBC}.

\section{Turbulence-based Load Alleviation Control}\label{sec:TBC}
\begin{figure*}
\centering
%	\centering
	\includegraphics[width=15cm]{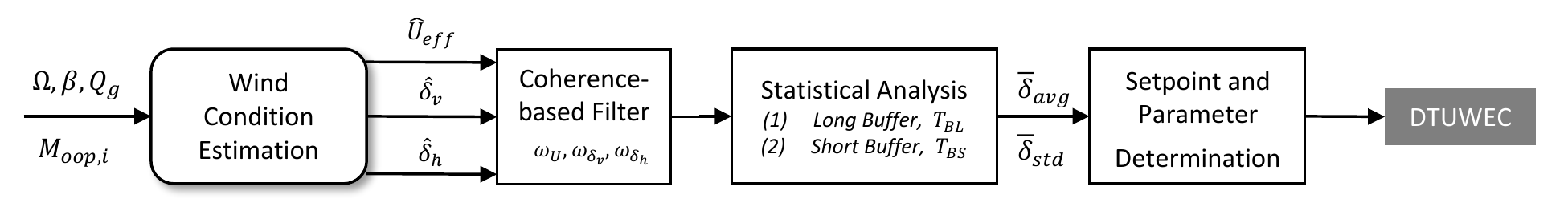}
	\caption{Conceptual diagram of turbulence-based load alleviation control.}
	\label{fig:ControlLogic}
\end{figure*}

The turbulence-based control concept is presented in \prettyref{fig:ControlLogic}, wherein the first feature block is the estimation of the rotor averaged wind condition as described in \prettyref{sec:RAWCEstimation}. The next three blocks contain the control logic for identifying the extreme wind conditions and determining the control parameters, including coherence-based filter, statistical analysis and set points or parameters determination.

\subsection{Controller Design}
The estimated wind information should be filtered by the second-order low-pass filter to only keep the highly correlated information, so the optimal cutoff frequency is chosen to be the frequency around 0.5 coherence as in \prettyref{fig:Coherence_WSE}. 
%The second order low pass filter in the Laplace domain,
%\begin{equation}
%    G_{filt}(s) = \frac{w_n^2}{s^2+2\zeta{w_n}s+w_n^2}
%\end{equation}
%where $w_n$ is the frequency and $\zeta$ is damping ratio, the default is 0.7.

The statistical quantities $\bar{\delta}_{avg}$ or $\bar{\delta}_{std}$ in a buffer are defined as:
\begin{equation}
\begin{aligned}
\bar{\delta}_{avg} = \frac{1}{N_L}\sum_{k-{N_L}}^{k}{\delta(t)}, \quad
\bar{\delta}_{std} = \sqrt{\frac{1}{N_L}\sum_{k-{N_L}}^{k}{(\delta(t)-\bar{\delta}_{avg})^2}} .\\
\end{aligned}
\end{equation}
where $k$ is the current discrete time step and $N_L$ is the total samples in each time averaging buffer. Two different lengths of the buffers are described as follows.

\subsubsection{For short-term extreme load purpose}
To prevent the extreme load induced by the extreme wind shear events, the short-term buffer $T_{BS} = 60$ s is proposed. While the wind speed $U\in[U_\mathrm{LB}, U_\mathrm{UB}]$, the percentage of power down-regulation $P_\mathrm{sp}$ is calculated as:
\begin{equation}
        \begin{aligned}
        P_\mathrm{sp}(X) &= 
                \left\{
                    \begin{aligned}
                         1 - \frac{1 - p_\mathrm{lim}}{\delta_\mathrm{UB} - \delta_{t,X}}(X - \delta_{t,X}),\qquad&\delta_{t,X}\leq{X}\leq\delta_\mathrm{UB},\\
                         1,\qquad&X < \delta_{t,X},\\
                         p_\mathrm{lim},\qquad&X > \delta_\mathrm{UB}.\\
                    \end{aligned}
                \right.
%        p &= \frac{1 - p_\mathrm{lim}}{\delta_\mathrm{UB} - \delta_\mathrm{thres}}(X - \delta_\mathrm{thres})
        \end{aligned}
\end{equation}
where $X$ represents either $\bar{\delta}_{avg}$ or $\bar{\delta}_{std}$, $\delta_\mathrm{UB}$ is the upper bound, the value 1 represents the power set point equals to rated power, while the value $p_\mathrm{lim}$ denotes the maximum derating percentage. Wind-dependent thresholds $\delta_{t,X}$ are chosen based on following guidelines: a) minimizing the influence on power production during normal operation in NTM cases; b) reducing the loads due to the extreme turbulence in ETM.

The $P_\mathrm{sp}$ sent to DTUWEC is determined by the same degree of importance of $\bar{\delta}_{avg}$ and $\bar{\delta}_{std}$ in \prettyref{eq:Psp}, which could be improved in the future work:
\begin{equation}\label{eq:Psp}
P_\mathrm{sp} = \mathrm{min}(P_\mathrm{sp}(\bar{\delta}_{ave}),P_\mathrm{sp}(\bar{\delta}_{std})) .\\
\end{equation}

The down-regulation strategies are described in \cite{HouLio2020EffectiveDown-regulation}, including both torque-based and rotor-speed-based down-regulation strategies. For the sake of brevity, the methods of implementation are not presented here.
%\begin{equation}
%    J = \int_{0}^{T_{BL}}(\alpha_{Load}M(U_{eff},TI,\delta) + \alpha_{Ele}P_{loss}(P_{sp})dt
%\end{equation}
%where $M$ is the surrogate model to represent component loads as function of the wind condition, 

\subsubsection{For long-term fatigue load purpose}
A long-term buffer $T_{BL}$ is required to represent the relatively long-term wind condition, which might be caused by long-term extreme weather. For instance, the wind condition is collected in each $T_{BL} = 10$ min buffer for several hours. If the turbine is constantly operating in extreme wind condition, down-regulation mode should be enabled for a longer period, avoiding the frequent switching between rated and derated power mode as in the control logic with short-term buffer.
 
Furthermore, the Proportional-Integral (PI) gains of conventional pitch regulated controller are gain-scheduled by pitch angle or wind speed~\cite{Lio2019AnalysisDown-regulation}, which can be tuned more aggressively during the long-term extreme turbulence event, in case of avoiding over-speeding and reducing load cycle magnitude. Thus, the PI gain $K_\mathrm{p}$ and $K_\mathrm{i}$ can be scheduled according to $\mathbf{K}$ function:
\begin{equation}
        [K_\mathrm{p},K_\mathrm{i}] = \mathbf{K}(P_\mathrm{sp},U_\mathrm{eff},\delta).
\end{equation}
The stability and robustness of different combination of control feature are not considered, which remain topics of future work.

\subsection{Performance Evaluation}\label{sec:Simulation}
The performance of the TLAC algorithm is analyzed by performing the simulation in DLC 1.2 and DLC 1.3. In order to achieve the optimal trade-off between power production and extreme load reduction, the threshold $\delta_\mathrm{thres}$ is chosen to be the 80\% quantile of the corresponding variable (dashed line in \prettyref{fig:Relationship_ETMNTM}). It can be observed that the thresholds lie above the $\bar{\delta}_{avg}$ and $\bar{\delta}_{std}$ for most of the NTM cases, which are as expected for minimizing the power loss during the normal operating condition. If further load reduction is required, the TLAC can be triggered occasionally in DLC 1.2 by reducing the threshold, then a certain amount of power production will be sacrificed. 

\begin{figure*}
\centering
	\centering
	\includegraphics[width=17cm,trim={0cm 0cm 0 0.5cm},clip]{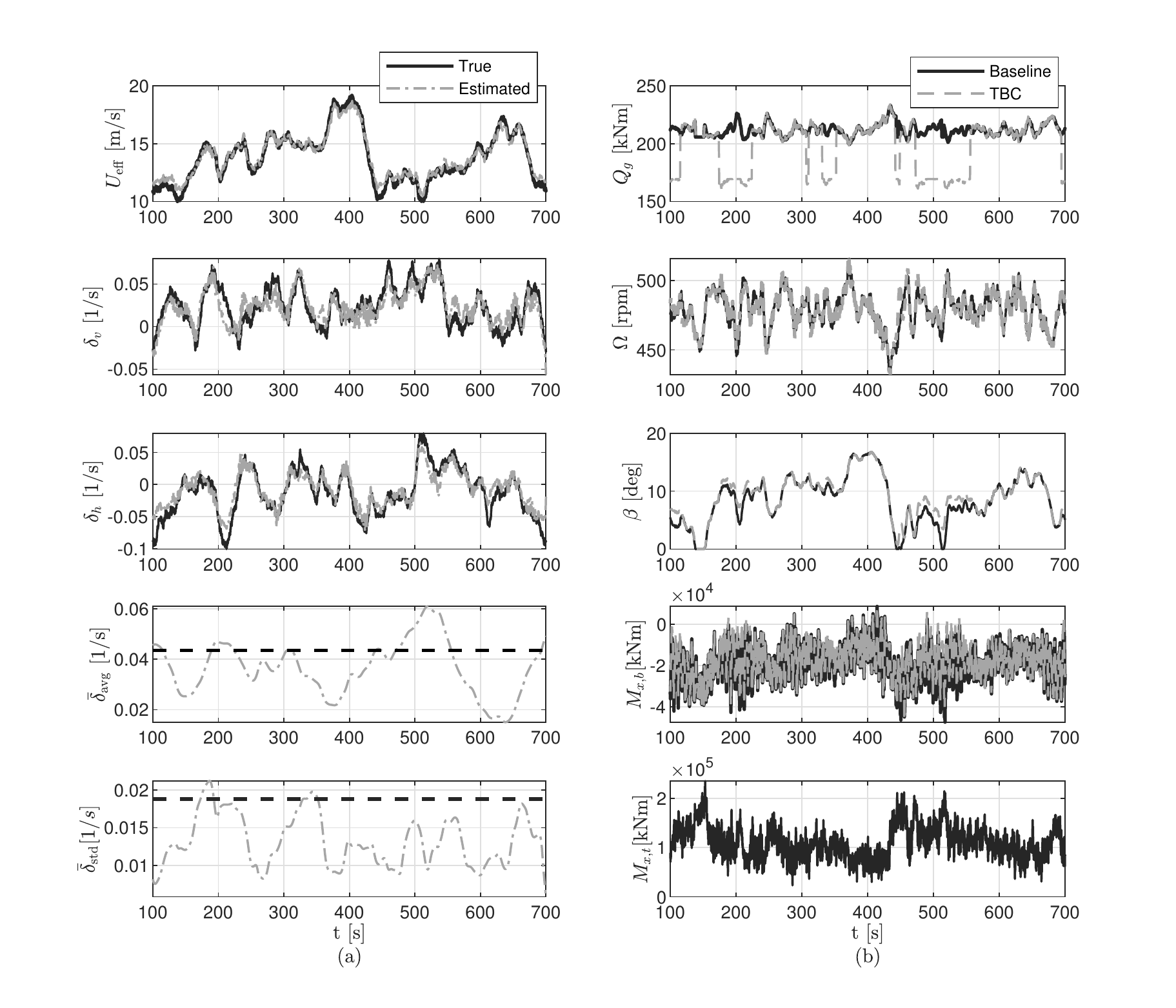}
	\caption{Comparison of baseline and TLAC. The ETM at a mean wind speed of 14 m/s. \\(a) Wind conditions. From top to bottom: rotor effective wind speed $U_\mathrm{eff}$, vertical shear $\delta_v$, horizontal shear $\delta_h$, the averaged resultant shear $\bar{\delta}_\mathrm{avg}$, and standard deviation of resultant shear $\bar{\delta}_\mathrm{std}$. The dashed line represents the threshold. \\(b) Wind turbine states. From top to bottom: generator torque $Q_g$, generator speed $\Omega$, pitch angle $\beta$, blade root flapwise bending moment $M_{x,b}$, tower bottom fore-aft bending moment $M_\mathrm{x,t}$.}
	\label{fig:Timeseries_14s03}
\end{figure*}

The effectiveness of TLAC during the extreme turbulence condition is clearly demonstrated in \prettyref{fig:Timeseries_14s03}. It demonstrates that the general trend of the wind condition can be captured by the proposed estimation method in \prettyref{fig:Timeseries_14s03} (a). The TLAC is triggered quite a lot in this ETM case as seen in \prettyref{fig:Timeseries_14s03} (b). While the TLAC is triggered, the turbine power is down-regulated to 80\%, thus several extreme load peaks have been mitigated around 200 s and 500 s. At some time steps, the extreme events are not well captured by the estimated wind condition, as seen around 150 s. The TLAC is seldom triggered during normal power production with the current threshold settings, in this way, the load reduction is achieved without the loss of annual energy production.

%\prettyref{fig:Timeseries_14s05} illustrates that a more stable generator speed can be achieved and the variation of tower top (TT) thrust $F_y$ can be reduced, resulting from the turbulence-based PI tuning.
%\begin{figure}
%\centering
%	\centering
%	\includegraphics[width=15cm]{Timeseries_14s05.eps}
%	\caption{Illustration of turbulence-based PI tuning.}
%	\label{fig:Timeseries_14s05}
%\end{figure}

\begin{figure*}
\centering
%	\centering
	\includegraphics[width=17cm,trim={0cm 0cm 0 0cm},clip]{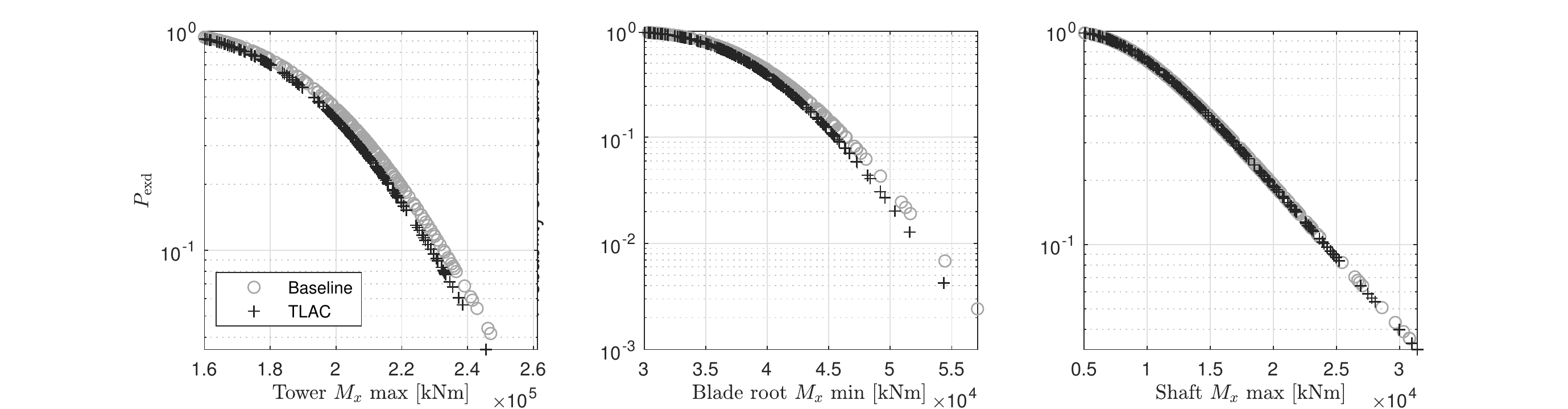}
	\caption{Exceedance probability with baseline and TLAC. From left to right: tower bottom fore-aft bending moment, blade root flapwise bending moment, and stationary shaft bending moment.}
	\label{fig:ExLoad_Compare}
\end{figure*}

The TLAC can affect the shape of the exceedance probability $P_\mathrm{exd}$ of load distribution, as  plotted in \prettyref{fig:ExLoad_Compare}, including the extreme loads of tower, blade and hub. The obvious reduction can be seen in the tower bottom fore-aft bending moment and the blade root flapwise bending moment. However, the TLAC has little impact on the shaft imbalance moment. This result may be explained by the fact that power down-regulation will reduce the thrust but has little influence on imbalance loads. Overall, the higher structural reliability can be achieved by this load reduction effect.

\section{Conclusion}\label{sec:conclusion}
The aim of the present research is to develop a feasible extreme turbulence indicator, which can be integrated into an adaptive controller to provide extreme load reduction benefits in extreme wind conditions. The results have confirmed that TLAC can effectively decrease the exceedance probability of extreme loads. Thus, the method is feasible to improve the structural reliability of a wind turbine.

The scope of this work was limited in terms of shear-based derating control, continued efforts are needed to explore the method to adapt the controller to the various wind condition, and to achieve the optimal trade-off between energy production and components loads. A further study with more focus on the potential of using LiDAR to obtain the measurement of approaching extreme wind condition is therefore suggested.

\bibliographystyle{IEEEtran}
\bibliography{references}
\end{document}